\documentclass[pre,preprintnumbers,amsmath,amssymb,twocolumn]{revtex4}
\usepackage{graphicx}
\usepackage{bm,color}
\usepackage{tikz}

\definecolor{labelkey}{cmyk}{.4,.2,0,0}

\newcommand{\nn}{\nonumber}

\begin{document}

\title{Electron and nuclear spin dynamics in the thermal mixing model of dynamic nuclear polarization}
\author{\bf  Sonia Colombo Serra$^1$, Alberto
Rosso$^{2}$ and Fabio Tedoldi$^1$}

\affiliation{\medskip
$^{1}$Centro Ricerche Bracco, Bracco Imaging Spa, via Ribes 5, 10010 Colleretto Giacosa (TO), Italy. \\
$^{2}$Universit\'e Paris-Sud, CNRS, LPTMS, UMR 8626, Orsay F-91405, France.\smallskip
}

\begin{abstract}
A novel mathematical treatment is proposed for computing the time evolution of dynamic nuclear polarization processes in the low temperature thermal mixing regime.
Without assuming  any {\em a priori} analytical form for the electron polarization, our approach provides a quantitative picture of the steady state that agrees with the well known Borghini prediction based on thermodynamic arguments, as long as the electrons-nuclei transition rates are fast compared to the other relevant 
time scales. Substantially different final polarization levels are achieved instead when the latter assumption is relaxed in the presence of a nuclear leakage term, 
even though very weak, suggesting a possible explanation for the deviation between the measured steady state polarizations and the Borghini prediction.
The proposed methodology also allows to calculate nuclear polarization and relaxation times, once the electrons/nuclei concentration ratio and the typical rates 
of the microscopic processes involving the two spin species are specified. Numerical results are shown to account for the manifold dynamical behaviours of typical DNP samples. 
\end{abstract}

\maketitle
 
\section{Introduction}
\label{intro}
Dynamic Nuclear Polarization (DNP) techniques are attracting increasing interest due to their proven ability
of improving the sensitivity of Nuclear Magnetic Resonance (NMR) experiments by several orders of magnitude, not only in the solid state \cite{vecchioartDNP1} but also in solution \cite{PNAS JHAL}.
Although the disruptive potential of DNP is nowadays well known and accepted, the articulated physical scenario underlying these complex phenomena is still puzzling the scientific community.

In the solid state, three different DNP regimes can be specified according to the resonance frequency $\omega_n$ of the nuclei to be polarized, the width $\Delta \omega_e$ of the Electron Spin Resonance 
(ESR) line of the free radicals used as polarization source and the magnitude of dipolar coupling between those radicals $1/T_{2e}$. 

When  $\omega_n>\Delta \omega_e$ the main mechanism for nuclei to polarize, known as Solid Effect (SE), proceeds via microwave (MW) assisted forbidden transitions involving simultaneous flip-flops or flip-flips 
of one electron and one nucleus in mutual dipolar interaction. The SE model has been extensively described in \cite{Abragam e Goldman} and the role played by spin diffusion was discussed in \cite{khut}. More recently, an exact quantum mechanical treatment of the SE based on the density matrix approach has been proposed by Hovav et al. \cite{Vega1, Vega2} and by Karabanov and collaborators \cite{kock}. 

On the other side, when $\omega_n<\Delta \omega_e$, nuclei can flip between different Zeeman levels within an energy conserving three particle mechanism involving a simultaneous flip-flop of two electron spins
(referred as $ISS$ process hereafter). Such process, driven by electron-electron and electron-nucleus time dependent dipolar interactions, does not involve forbidden transitions and it is thus in general more 
effective than SE. As far as the typical interaction time $T_{2e}$ between different paramagnetic centres is long compared to the electron spin lattice relaxation time $T_{1e}$ ({\em i.e.} the unpaired electrons are - on average - relatively far from each other), the polarization mechanism is referred as Cross Effect (CE) and was first analyzed in \cite{k1, k2, HH1, HH2}. The CE model has been successfully exploited to
describe DNP with bi-radicals \cite{Griffin1, Griffin2}, {\em i.e.} polarization procedures where the polarizing agents are tailored molecules carrying two unpaired electrons having resonance frequencies differing 
exactly by $\omega_n$. Similarly to SE, also the CE model has been  rigorously computed by $ab$ $initio$ quantum mechanical techniques \cite{Vega3, Vega4}.

However, the novel applications which in the last decade have renewed the attention on DNP due to their potential impact on biosciences \cite{PNAS, CR}, are based on samples which generally fall in none 
of the two models previously described. The nuclei of interest for this kind of applications  (typically $^{13}$C or $^{15}$N)  have a low-gyromagnetic ratio $\gamma_n$ and, as a consequence, a resonance 
frequency $\omega_n$ significantly smaller than the width of the typical radicals used as polarizing agent (trityls or nitroxides). Moreover, the concentration of these radicals, usually above 10 mM in the 
solution to polarize, leads to non negligible electron spin-spin interactions, {\em i.e.} to the condition $1/T_{2e} >> 1/T_{1e}$ that define the Thermal Mixing (TM) regime. Under TM assumptions, when a 
transition is saturated by an external, frequency selective, microwave field, the whole electron spin distribution reacts through energy conserving flip-flops (spectral diffusion) and evolves 
towards a new steady state different from Boltzman equilibrium. Hence, via the previously introduced $ISS$ processes, the nuclear Zeeman populations are also perturbed and, depending on the frequency of the saturated electron
transition, possibly result in depleting or enhancing the population of the ground state and thus in an enhanced nuclear magnetic order. At high temperature, where a linear expansion of the density matrix can be used, 
the evolution of the electron and nuclear polarizations is accounted by a set of rate equations formulated by Provotorov \cite{Provotorov1, Provotorov2, Provotorov3, Provotorov4, Abragam e Goldman}.
At the typical temperatures (about 1 K) where most of the DNP experiments aimed to obtain substantial nuclear polarization values ($>$ 0.1) are performed, however, a linear expansion of the density 
matrix is not  allowed and consequently the Provotorov approach does not apply. Fortunately in these conditions, the electron resonance lines of the free radicals used as polarizing agents are 
normally inhomogeneously broadened and can likely be depicted as the convolution of several individual packets of given resonance frequency, mutually connected by the electron-electron dipolar 
interaction \cite{Abragam e Goldman}.  Hyperpolarization via TM at low temperature has been first discussed by Borghini,   
who also calculated the steady state solution for the electron and nuclear polarization after imposing certain constrains to the model \cite{Borghini PRL}. The same - steady state - result has been 
achieved by Abragam and Goldman \cite{Abragam e Goldman} using a slightly different mathematical procedure. The Borghini prediction has the merit of reproducing, at least qualitatively, many observations 
obtained in DNP experiments at low temperature \cite{JHAL2008, Sami, Texas}. It has, at the same time, several limitations, the major of them listed below.
\begin{itemize}
\item[$i$)] It provides a picture of the steady state polarization without contributing in any way to the understanding of the time evolution of nuclear spin order towards equilibrium.
\item[$ii$)] It is obtained by assuming an {\em a priori} analytical form for the electron spin polarization. 
\item[$iii$)] It is derived under strong saturation conditions and in the limit of perfect contact between the electron and the nuclear reservoirs; the latter hypothesis consists in assuming highly effective $ISS$ processes.
\end{itemize}
As a consequence of the constrains ii) and iii), the model leads to a final steady state nuclear polarization which substantially depends only on the lattice temperature, the magnetic field in which the DNP phenomenon takes place and 
the width of the electron resonance line. The sample specific parameters like electron and nuclear spin-lattice relaxation rates and the relative concentration of the two spin species play only a secondary role. 

In this work we aim to overcome these three limitations by introducing a dynamic analysis of the low temperature TM model (described in detail in Section \ref{MandM})
based on a set of rate equations presented in Section \ref{Rateequationsapproach} which spontaneously provide the time evolution laws of the electron and nuclear polarizations
without any {\em a priori} assumption on their functional form.
The results obtained by numerically solving the rate equations are reported in Section \ref{NumericalResults} and discussed in Section \ref{Discussionandconclusion}, with particular emphasis
on the modifications occurring in the steady state and in the dynamic parameters in the presence of nuclear leakage and finite electron-nucleus exchange.
The technical arguments underlying the derivation of the rate equations set are given in Appendix \ref{DetailedBalance} and \ref{RateEqDerivation}, while in Appendix \ref{Borghirel}, for convenience of the
reader, the derivation of the Borghini equation for the steady state as proposed by Abragam and Goldman \cite{Abragam e Goldman} is briefly recalled.

\section{Model description}
\label{MandM}
A system made up of $N_n$ nuclear spins $\textbf{I}$ with Larmor frequency $\omega_n$ and $N_e$ electron spins $\textbf{S}$ with mean Larmor frequency $\omega_e$ is considered ($S=I=1/2$). Typically  $\omega_e$ is about three orders of magnitude higher than $\omega_n$. Since all terms in the nuclear Hamiltonian with the exception of the Zeeman one are small in comparison to the other energy scales considered, nuclei are assumed to resonate all at the same frequency.

\begin{figure}[t]
 \includegraphics[width=8.6cm]{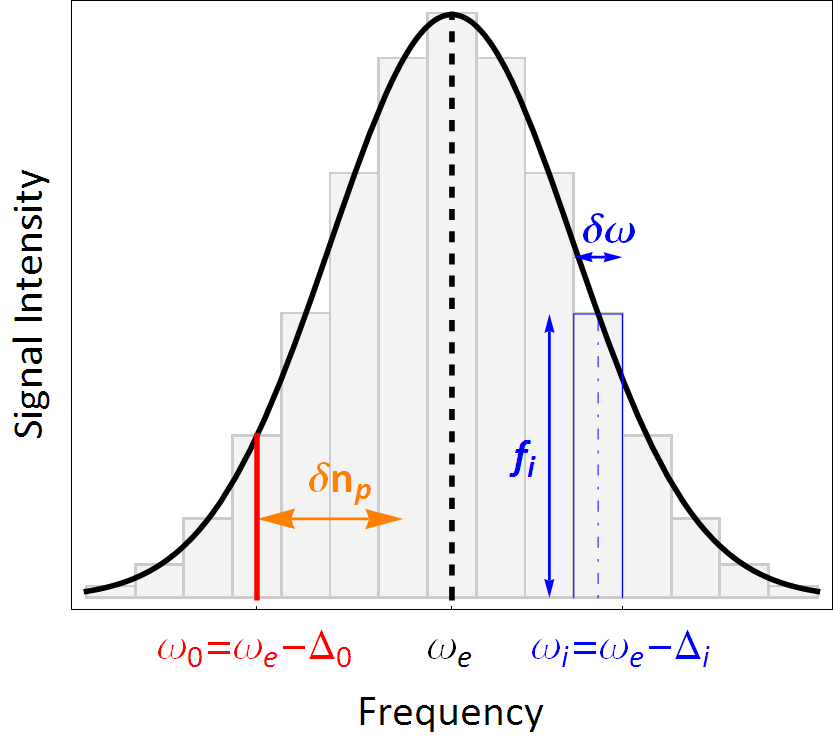}
\caption{Discretization of the ESR line. Each basic packet centered at $\omega_i = \omega_e - \Delta_i$  is characterized by a width $\delta \omega$ and a weight $f_i$. By way of example
the irradiation frequency is  set to $\omega_0 = \omega_e - \Delta_0$ and the nuclear Larmor frequency corresponds to three basic packets ($\delta n_{p}=3$).}
\label{figureESR}
\end{figure}

The first assumption of the model is that the main contribution to the ESR line shape is  the spread of $g$-factors (inhomogeneous broadening of the I type), that reflects the different orientation of the 
unpaired magnetic moments with respect to the external magnetic field due to single ion anisotropy. The practical relevance of this assumption is confirmed, for instance, by
the experimental evidence obtained from the widely used trityl radical family \cite{JHAL2008}. Inhomogeneously broadened lines are conveniently decomposed in a sequence
of narrow individual spin packets of frequency $\omega_i = \omega_e - \Delta_i$, width $\delta \omega$ and relative weight $f_i$ (see \figurename~\ref{figureESR}) such that: 
\begin{eqnarray}
\label{fideltai}
\sum_{i} f_i = 1 \\ 
\sum_{i} f_i \Delta_i = 0,  \nonumber
\end{eqnarray}
valid for any ESR line.
The continuous limit is recovered when $\delta \omega \rightarrow 0$.
For each electron packet, a local polarization $P_{ei} = 2 \left\langle S^{i}_{z}\right\rangle$ may  be defined, where $\left\langle \right\rangle$ stands both for the quantum mechanic 
expectation value and the average over all electron spins belonging to the $i$-th packet.
The second assumption of the model is that $\omega_n$ is smaller than the ESR line width, as actually is the case for low $\gamma_n$ nuclei with almost any type of radical or for
high gamma nuclei, such as $^1$H, with broad line radicals ({\em e.g.} nitroxides). 
In view of the mathematical description proposed in the next section, it is useful to introduce at this point the variable
\begin{equation} 
\delta n_p = \frac{\omega_n}{\delta \omega},
\label{deltanp}
\end{equation}
{\em i.e.} the number of electron packets corresponding to the nuclear Larmor frequency (see \figurename~\ref{figureESR}). 

\begin{figure}[b]
 \includegraphics[width=8.6cm]{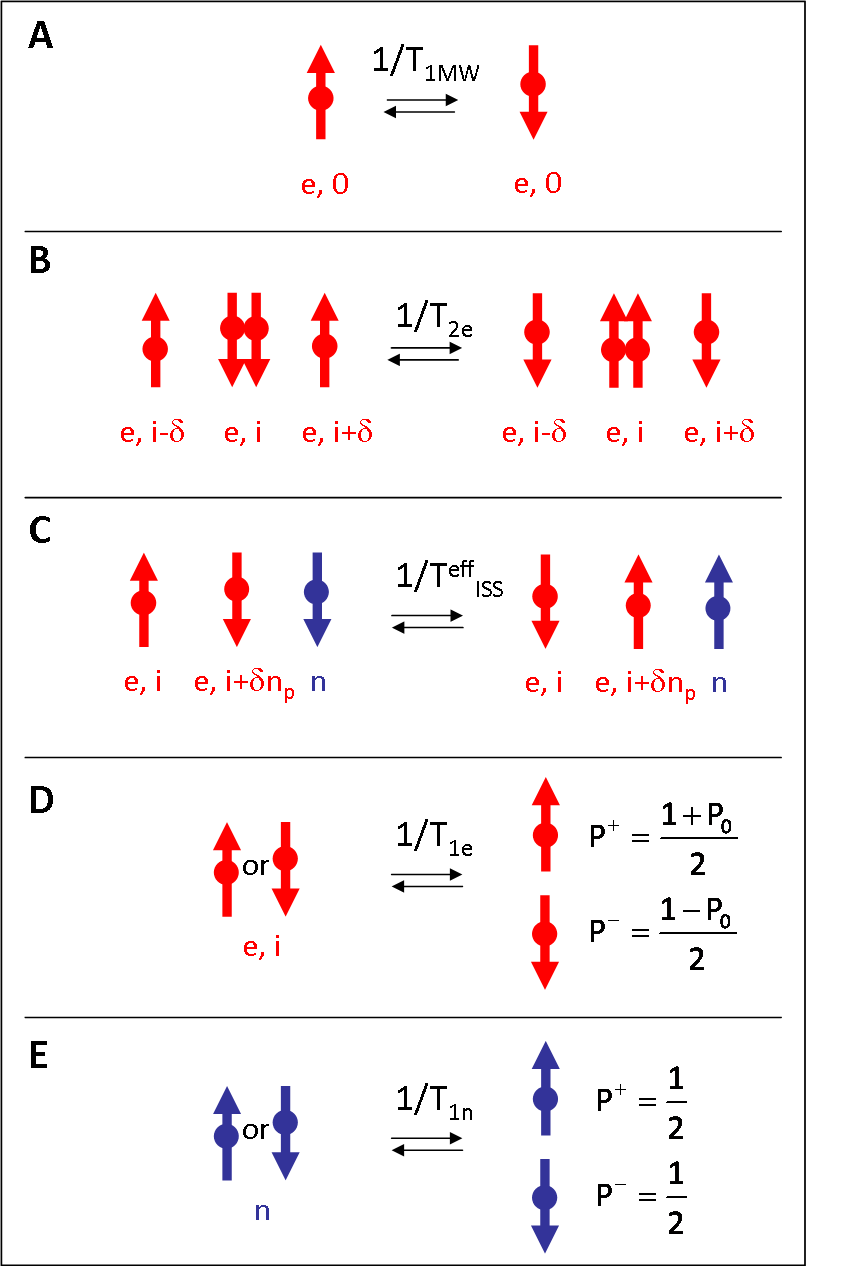}
\caption{Microscopic events in the thermal mixing model. Electron spins belonging to a generic packet $i$ are represented in red with a subscript $e, i$, nuclear spins in blue with a subscript $n$. Panel A: electron flip due to MW irradiation; Panel B: spectral diffusion ($\delta$ is a generic number of packets); Panel C: $ISS$ processes; Panel D: electron spin-lattice relaxation; Panel E: nuclear spin-lattice relaxation (``leakage''), under the assumption $P_{0,n}\approx0$ (see Eq.(\ref{P0n})).}
\label{figureTransitions}
\end{figure}

The third assumption is that the spin dynamics of the system is governed only by the five processes depicted in \figurename~\ref{figureTransitions} and briefly explained here below.

\textsl{Microwave irradiation.}
Single electron transitions are stimulated by an external MW field at $\omega_0 = \omega_e - \Delta_0$ that leads the electron spin system out of equilibrium (panel A of \figurename~\ref{figureTransitions}). 
The characteristic time of this process is named $T_{1 \text{MW}}$. For high microwave power $T_{1\text{MW}} \rightarrow 0$ and the packet $\Delta_0$ is saturated.

\textsl{Spectral diffusion.}
The basic transition of this process, which involves only electrons and conserves both the total energy and the total electron polarization,  is represented in \figurename~\ref{figureTransitions}, panel B. 
Spectral diffusion transitions are promoted by the dipolar interaction among electrons and are characterized by a time constant $T_{2e}$ which in the thermal mixing regime is assumed to be  much shorter 
than any other relevant time scale (in practice, in a solid solution used for DNP, it is typically $<1$ $\mu$s \cite{JHAL2008}).

\textsl{$ISS$ process.}
The mechanism is sketched in \figurename~\ref{figureTransitions}, panel C, and consists in a simultaneous flip-flop of two electron spins, belonging to packets separated by $\omega_n$, compensated by a nuclear spin flip. This process allows an energy transfer between nuclei and electrons and while the total energy and the electron polarization are conserved, the nuclear polarization is not. 
In a real system the $ISS$ process, being promoted by dipolar interactions, involves only nuclei which are sufficiently close to paramagnetic centres. Thanks to nuclear spin diffusion, the local information is then spread throughout the entire system.
A model which takes into account both these aspects (local dipolar interaction and spatial diffusion) would be very cumbersome to treat, because of the $N_n$ degrees of freedom. 
In the present work we adopt a mean field approach where all possible terns comprising a nucleus and two electrons flip with a characteristic time $T_{\text{ISS}}^{\text{eff}}$ which does not depend on the mutual distances between the three particles. Setting low values of $T_{\text{ISS}}$ the model mimics the fast spin diffusion limit, while setting high values of $T_{\text{ISS}}$ the model mimics the slow spin diffusion limit.

\textsl{Electron spin-lattice relaxation.}
The process, shown in \figurename~\ref{figureTransitions}, panel D, accounts for the contact between the electron system and the lattice which leads $P_{ei}(t)$ towards the electron thermal equilibrium polarization $P_0$:
\begin{equation}
P_0 = -\tanh \frac{\hbar \omega_e}{2 k_B T},
\label{P0e}
\end{equation}
where $T$ is the temperature of the lattice.
Its characteristic time constant $T_{1e}$ ranges from hundreds of milliseconds to few seconds \cite{JHAL2008, Zhou}, for T $\lesssim 10$ K.

\textsl{Nuclear spin-lattice relaxation (``leakage'').}
The process is shown in \figurename~\ref{figureTransitions}, panel E. The nuclear system is directly in contact with the lattice via slow processes (where electrons are not involved), with a characteristic time $T_{1n} > 10^3$ s, 
also called ``leakage'' terms. Analogously to the latter case, these processes lead nuclear polarization, $P_{n}(t)$, towards the  thermal equilibrium value $P_{0,n}$:
\begin{equation}
P_{0, n} = \tanh \frac{\hbar \omega_n}{2 k_B T} \approx 0.
\label{P0n}
\end{equation}

In the next section, the temporal evolution of the nuclear polarization for the model described above will be determined by a closed set of equations involving only $P_n(t)$ and $P_{e,i}(t)$.

\section{Rate equations approach}
\label{Rateequationsapproach}
To investigate the dynamic evolution of the model summarized in the previous section, we introduce a system of rate equations and solve them numerically.
The discrete time step $dt$ for the numerical integration is defined as the inverse of the sum of the single rates for the five processes  sketched in \figurename~\ref{figureTransitions}. 
 
\textsl{Microwave rate.} The rate of all possible microwave events is $W_{\text{MW}} = N_e f_0 / T_{1 \text{MW}}$. 

\textsl{$ISS$ process rate.} In the mean field approximation, the number of the possible processes involving two electrons and one nucleus is given by $\sum_i f_i N_e f_{i+\delta n_p} N_e N_n$. 
The total $ISS$ rate needs to linearly scale with the size of the system in order to assure a correct thermodynamic limit. To achieve this, as is usually done for fully connected models, the effective time constant of each three particle process must depend on the system size and scale as:
\begin{equation}
T_{\text{ISS}}^{\text{eff}}=T_{\text{ISS}}N_e N_n,
\end{equation}
where the constant $T_{\text{ISS}}$ is size independent. Then the total rate of all $ISS$ events writes $W_{\text{ISS}} = N_e \sum_i f_i f_{i+\delta n_p} / T_{\text{ISS}}$.

\textsl{Electron and nuclear spin-lattice relaxation rate.} The rate of all the electron spin-lattice relaxation processes is $W_{e} = N_e/T_{1e}$.
Similarly the rate of all the nuclear spin-lattice relaxation processes is $W_{n} = N_n/T_{1n}$. 

\textsl{Spectral diffusion rate.} Being proportional to $1/T_{2e}$, it is assumed to be by far the highest total transition rate among those characterizing the five considered mechanisms.
A large rate makes the time step $dt$ very small and consequently produces a dramatic slow-down of the simulation procedure.
To avoid that, the dynamic problem has been separated in two steps: a ``Short-Term Thermalization'' (STT) step involving spectral diffusion and a ``Long-Term Evolution'' (LTE) where 
the system evolves according to defined differential equations under the action of the remaining processes.
Depending on the value of $T_{1\text{MW}}$, the microwave pumping can be considered either as a fast process (contributing to STT) or as a player in the LTE.
In the first case the time step is defined as:
\begin{equation}
dt = \frac{1}{W_e + W_{\text{ISS}} + W_n},
\end{equation}
while in the second case one has:
\begin{equation}
dt = \frac{1}{W_e + W_{\text{ISS}} + W_n + W_{1\text{MW}}}.
\end{equation}

The numerical procedure can be summarized as follows:
\begin{equation}
\left\{\begin{array}{c}
	P_{e, i}^0(t) \\
	\\
	P_{n}(t)
\end{array} \right.
\stackrel{\text{STT}}{\Longrightarrow}  
\left\{\begin{array}{c}
	P_{e, i}(t) \\
	\\
	P_{n}(t)
\end{array}\right.
\stackrel{\text{LTE}}{\Longrightarrow} 
\left\{\begin{array}{c}
	P_{e, i}^0(t+dt) \\
	\\
	P_{n}(t+dt)
\end{array} \right. \nonumber
\label{Steps}
\end{equation}

\subsection{STT}
The goal of this step is to ``thermalize'' the profile of the electron polarization $P^0_{ei}(t)$ at a generic time t, under the action of those processes considered as ``fast''.
The steady state profile $P_{e, i}(t)$ can be obtained imposing the detailed balance condition (see Appendix \ref{DetailedBalance}):
\begin{equation}\label{tanh}
P_{e, i}(t) = -\tanh\left[ \beta (\Delta_i-c)\right].
\end{equation}
This equation depends on two parameters that can be computed using conservation principles. 
In this respect we need to discuss separately the two cases where spectral diffusion only or both spectral diffusion and microwave promoted processes are viewed as fast.

\subsubsection{Spectral diffusion only.}
This case is relevant for relaxation experiments (with microwave field off by definition) or  when a non saturating microwave field is applied in a DNP experiment.
To calculate the two parameters $c$ and $\beta$, the conservation of {\em both} the total energy and the total polarization is imposed:
\begin{eqnarray}
&& \sum f_i \Delta_i \left(P_{e, i}(t)- P_{e, i}^0(t)\right)=0   \\
&& \sum  f_i \left(P_{e, i}(t)- P_{e, i}^0(t)\right)=0.  \label{due}
\end{eqnarray}

\subsubsection{Spectral diffusion and saturating microwaves.}
For high irradiation power, the microwaves act as an infinite bath for energy exchange and force the polarization $P_{e, 0}(t)$ to 0 and in turn, by Eq.(\ref{tanh}), $c= \Delta_0$.
In order to evaluate the second parameter $\beta$, the output of the STT step is conveniently written as:
\begin{eqnarray}
P_{e,i}(t)&=& P^0_{e, i}(t) + \delta P_{e, i}   \quad \mbox{if} \, i \ne 0   \\
P_{e, 0}(t)&=& P^0_{e, 0}(t) + \delta P_{e, 0} +\delta P^{MW}
\end{eqnarray}
where $\delta P_{e, i}$ and  $\delta P^{MW}$ are the variation of the polarization of the $i$-th packet induced by spectral diffusion and by microwaves respectively.
The conservation of the total energy and of the total polarization can now be written:
\begin{eqnarray}
&& \sum f_i \Delta_i \delta P_{e, i}=0   \\
&& \sum  f_i \delta P_{e, i}=0.  \label{dueb}
\end{eqnarray}
Eq.(\ref{dueb}) can be recast as $f_0 \delta P_{e, 0}=-\sum_{i \ne 0}  f_i  \delta P_{e, i}$, so that:
\begin{eqnarray}
\lefteqn{\sum f_i \Delta_i \delta P_{e, i}}  \nn \\
&=& \sum_{i \ne 0}  f_i   \Delta_i  \delta P_{e, i} + f_{0} \Delta_0 \delta P_{e, 0} \nn \\ 
&=& \sum  f_i  ( \Delta_i-\Delta_0)  \delta P_{e, i} =  0. 
\end{eqnarray}
Summing on both sides  $ \sum  f_i  ( \Delta_i-\Delta_0)  P^0_{e, i}(t)$ we get: 
\begin{equation}
 \sum f_i  ( \Delta_i-\Delta_0)   P_{e, i}(t) = \sum  f_i  ( \Delta_i-\Delta_0)  P^0_{e, i}(t).
\end{equation}
The condition for $\beta$
\begin{equation}
\label{ourbeta}
\sum f_i \left\{P^0_{e, i}(t)+\tanh \left[\beta(\Delta_i-\Delta_0)\right]\right\}(\Delta_i-\Delta_0) = 0
\end{equation}
identifies the unique solution for $P_{e, i}(t)$.

\subsection{LTT}
\label{RateEq}
In this second step the profile of the electron polarization $P^0_{e, i}(t+dt)$ at time  $t+dt$ is deduced from the output of the STT.
When microwaves are off (relaxation) or they have been already taken into account in the STT, the rate equations for $P_{e, i}$ and $P_n$,
as discussed in Appendix \ref{RateEqDerivation}, can be set down in the form:
\begin{eqnarray}
\label{rateeq}
 \lefteqn{P_{e, i}^0(t+ dt) = P_{e, i}(t)+} \nn \\
&+& dt \left( \frac{P_0-P_{e, i}(t)}{T_{1e}} + \frac{f_{i-\delta n_p} \Pi_- + f_{i+\delta n_p} \Pi_+ }{2 T_{\text{ISS}}} \right) \\
 \lefteqn{P_n (t+ dt)= P_n (t)+} \nn \\
&+& dt \left[ \frac{P_{0n}-P_n(t)}{T_{1n}}- \frac{N_e}{2 T_{\text{ISS}} N_n} \sum f_i f_{i+\delta n_p} \Pi_n  \right]  \nonumber
\end{eqnarray}
where $\Pi_-=\Pi_-(i,t), \Pi_+=\Pi_+(i,t), \Pi_n=\Pi_n(i,t)$ are given by the expressions:
\begin{eqnarray}
 \Pi_- &=& P_{e, i-\delta n_p}(t) \small{-}P_{e, i}(t)\small{+}P_n(t)\left[1\small{-}P_{e, i-\delta n_p}(t)P_{e, i}(t)\right] \nonumber \\ 
 \Pi_+ &=& P_{e, i+\delta n_p}(t) \small{-} P_{e, i}(t)\small{+}P_n(t)\left[1\small{-}P_{e, i+\delta n_p}(t)P_{e, i}(t)\right] \nonumber \\
 \Pi_n &=& P_{e, i+\delta n_p}(t)\small{-}P_{e, i}(t)\small{+}P_n(t)\left[1\small{-}P_{e, i+\delta n_p}(t)P_{e, i}(t)\right]. \nonumber
\end{eqnarray}
The effect of a non saturating MW irradiation can be included by the term $-P_{e,0}(t)/T_{1 \text{MW}}$ in the rate equation for $P_{e, 0}(t)$, describing the electron polarization of the irradiated packet.

\section{Numerical Results}
\label{NumericalResults}
In this section we exploit the formalism described above to calculate, in specific situations, those curves which are normally measured 
in DNP experiments, namely build-up and relaxation curves. The so-called build-up curve describes the polarization growth over time under 
selective microwave irradiation and it is characterized by two parameters, the final nuclear polarization $P_n$:
\begin{equation} 
P_n = P_n(t\rightarrow\infty),
\end{equation}
and the characteristic time $T_{\text{pol}}$ that we define by the equation:
\begin{equation}
\label{Tpoleq}
P_n(t=T_{\text{pol}})=P_n \left(1- \frac{1}{e}\right),
\end{equation}
with $P_n(t=0)=0$.  $T_{\text{pol}}$ corresponds to the usual time constant in case of exponential build-up but, through Eq.(\ref{Tpoleq}), one has a more general quantification of the growing speed regardless the specific form of the polarization function. 

Relaxation curves, on the other hand, represent the spontaneous (at microwaves off) equilibrium recovery of a system that is prepared out of equilibrium at $t=0$.
The characteristic time is named $T_{\text{relax}}$ and, if one starts with a fully polarized system ($P_n(t=0)=1$), the constant $T_{\text{relax}}$ is defined by:
\begin{equation}
P_n(t=T_{\text{relax}})= P_{0, n} \left(1- \frac{1}{e}\right)+\frac{1}{e},
\label{Trelax}
\end{equation}
where $P_{0, n}$ is defined in Eq.(\ref{P0n}).

\textsl{Setting of the parameters.} In order to make the results of our simulation relevant for a deeper understanding of experimental data, a number of basic
parameters are borrowed from some of the best known DNP samples for biomedical applications. In this framework, the typical number of  $^{13}$C nuclei 
(that in most cases are the object of DNP) is in the range $5$ - $15$ M,  while the radical concentration is $10$ - $20$ mM \cite{JHAL2010}. Is thus reasonable, for our purposes,
to set $N_n/N_e = 1000$. Concerning the distribution of the electron spin resonances, it looks worth to use a Gaussian function with a full width at half maximum $\Delta \omega_e = 63$ MHz, as surrogate of the ESR line of the trityl doped samples studied in \cite{JHAL2008}. This is actually one of the few ESR line widths which have been measured in a 
standard DNP environment (a magnetic field of $3.35$ T and a temperature of $1.2$ K) and that turned out to be inhomogeneously broadened by $g$-factor spreading.
Finally, despite the fact that our approach is not restricted to the case of saturating microwaves, we work under the approximation $T_{1MW} \rightarrow\infty$ 
to better mimic what is normally done in actual experiments.

\textsl{ESR line discretization.} The Gaussian function used as model of the ESR line is truncated at $3 \sigma$, where $\sigma = 27$ MHz. This defines a frequency interval of about $160$ MHz
to be covered by electron spin packets. Three different discretizations of the line are employed, with different packets widths: $\delta \omega = 32$, $10.6$ and $6.4$ MHz. 
The corresponding number of packets $N_p$ is thus equal to $5$, $15$ and $25$. The value of $\omega_n$ is set at $32$ MHz (resonance frequency of $^{13}$C 
nuclei at the typical DNP magnetic field of $3.35$ T) which corresponds respectively to $\delta n_p = 1, 3$ and $5$ for the three selected values of $N_p$. 
The MW frequency is set equal to $\omega_0 = \omega_e - \sigma$, which corresponds to saturating the packet $2$ (for $N_p = 5$), $5$ (for $N_p = 15$) 
and $8$ (for $N_p = 25$). For this value of $\omega_0$ the final nuclear polarization is maximal. 

The results we show are obtained by integrating the rate equations system (\ref{rateeq}) together with the STT step defined by the detailed balance condition (\ref{tanh}), 
with $c = \Delta_0$ and $\beta$ set by Eq.(\ref{ourbeta}). The boundary conditions of the ESR line shape are correctly implemented imposing 
$f_{i+\delta n_p} = 0$ ($f_{i-\delta n_p} = 0$) if $i+\delta n_p > Np$ ($i-\delta n_p <0$).

\begin{figure}[t]
 \includegraphics[width=8.6cm]{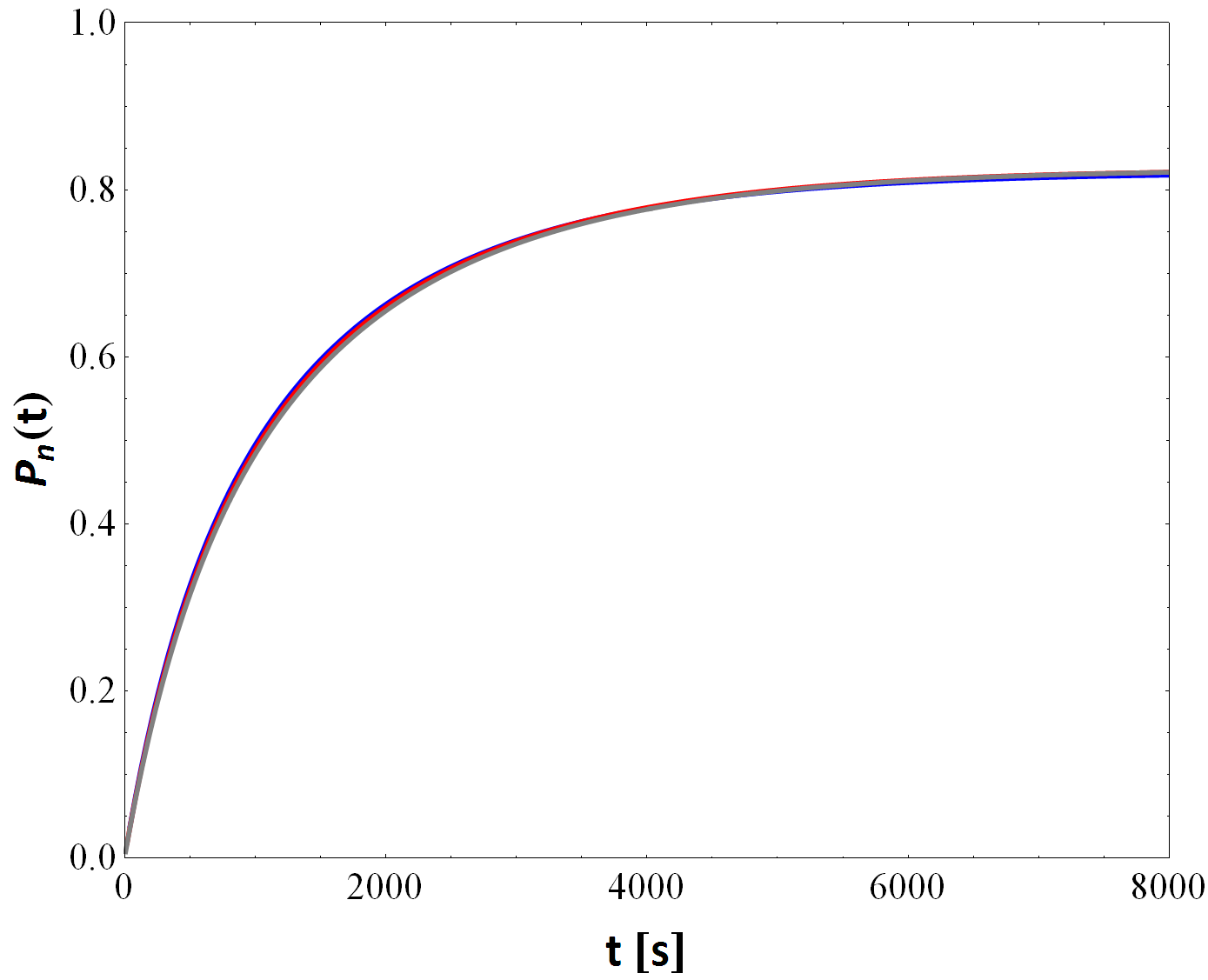}
\caption{Polarization build-up curves at 3.35 T and 1.2 K of a DNP system characterized by the following parameters:  $N_n/N_e = 1000$, $T_{1e} = 1$ s, $T_{\text{ISS}} = 0.001$ s, $T_{1n}\rightarrow\infty$, for different discretizations of
the ESR line: $N_p=5$ (blue line), $N_p=15$ (red line), $N_p=25$ (gray line). The final nuclear polarization is $P_n = 0.825$ reached with a characteristic time constant $T_{\text{pol}} = 1140$ s.}
\label{figure1}
\end{figure} 

\subsection{Fast $ISS$ limit}
The case of highly effective contact between electrons and nuclei, which corresponds to very fast $ISS$ processes compared to electron spin lattice relaxation
($T_{\text{ISS}}\ll T_{1e}$), is considered first, by setting $T_{\text{ISS}} = 10^{-3} T_{1e}$ and  $T_{1e} = 1$ s (the latter from \cite{JHAL2008}) in our computational tool.

In \figurename~\ref{figure1}, three different build-up curves, computed with an increasing value of $N_p$ and in absence of leakage ($T_{1n}\rightarrow\infty$) are presented. 
The continuum limit is approached very fast and a good convergence is reached already with $N_p = 5$. The final nuclear polarization is $P_n = 0.825$ and the characteristic time constant
of the polarization build-up curves $T_{\text{pol}}$ is equal to $1140$ s.

\begin{figure}[t]
 \includegraphics[width=8.6cm]{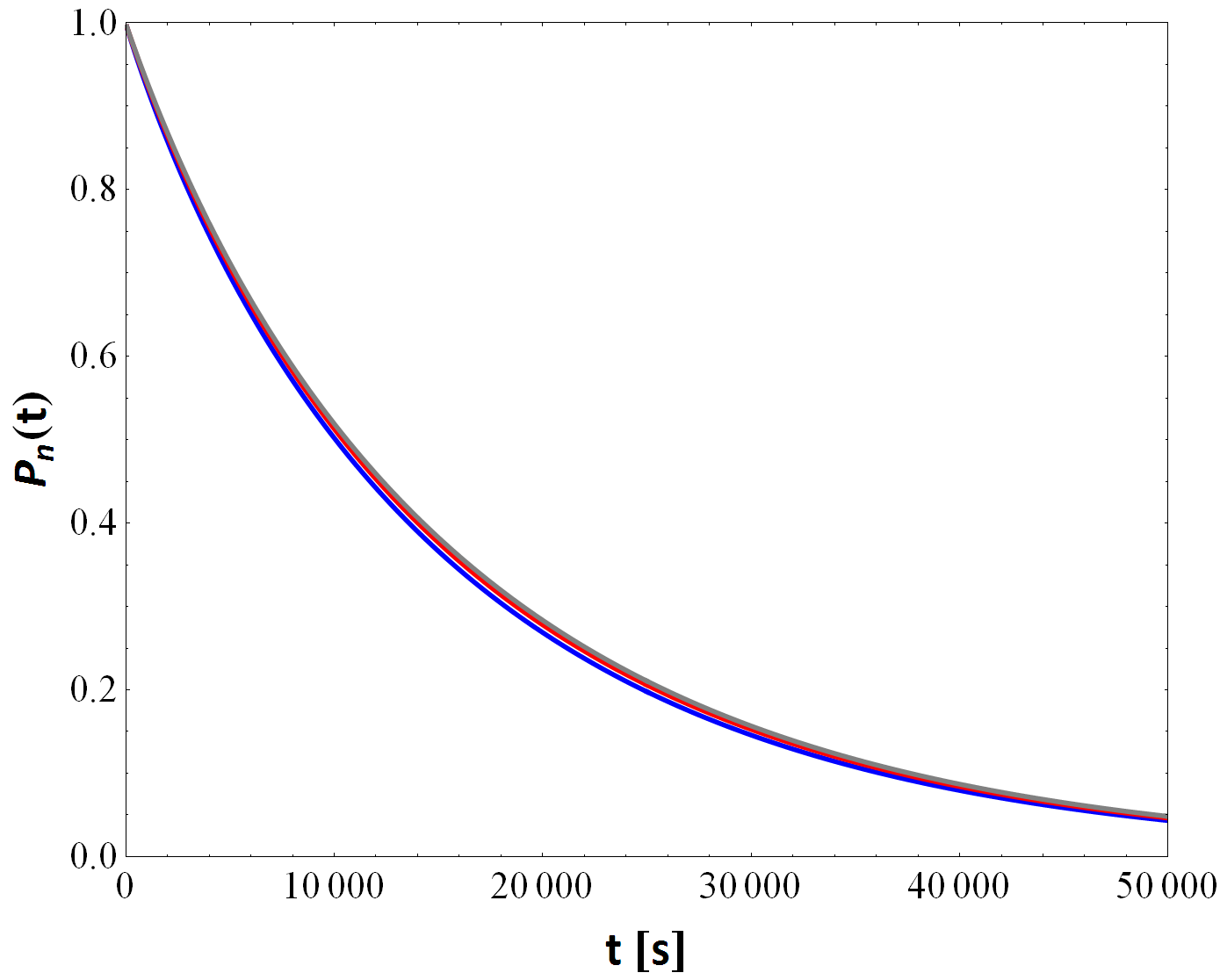}
\caption{Relaxation curves at 3.35 T and 1.2 K of a DNP system with $N_n/N_e = 1000$, $T_{1e} = 1$ s, $T_{\text{ISS}}$ = $0.001$ s, $T_{1n}\rightarrow\infty$, when $N_p=5$ (blue line), $N_p=15$ (red line), $N_p=25$ (gray line). The relaxation time is about $15600$ s.}
\label{figure2}
\end{figure}

In \figurename~\ref{figure2}, three relaxation curves, describing the evolution of the nuclear polarization in the absence both of microwaves and leakage and calculated with a different discretization
of the ESR line, are shown. Again, the curve with $N_p = 5$ is already representative of the continuous limit. The characteristic time $T_{\text{relax}}\approx15600$ s,  evaluated by means of 
Eq.(\ref{Trelax}) with $P_{0, n} = 0$, turns out to be about $10$ times longer than the corresponding $T_{\text{pol}}$.

\begin{figure}[b]
 \includegraphics[width=8.6cm]{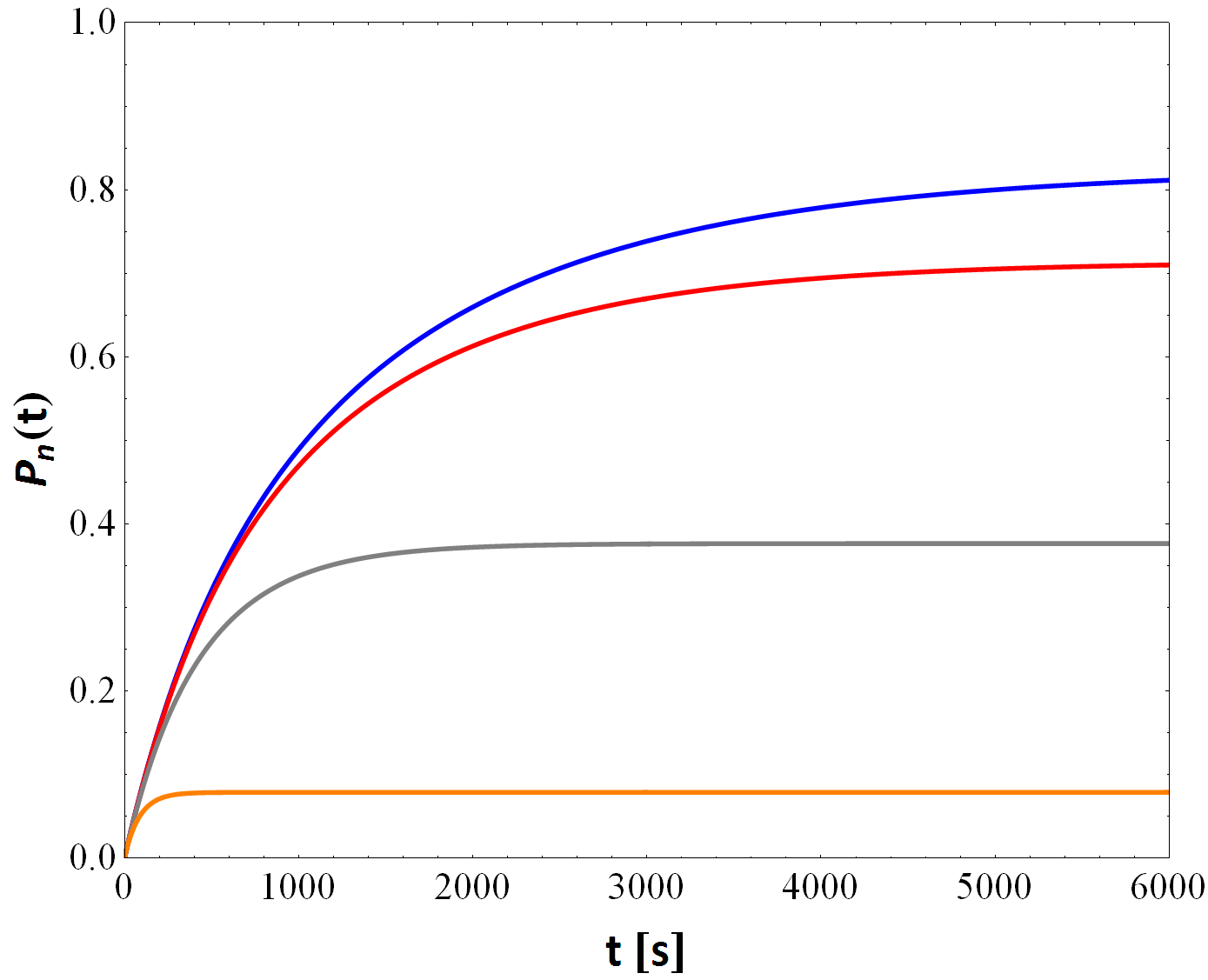}
\caption{Polarization build-up curves at 3.35 T and 1.2 K of a DNP system with $N_n/N_e = 1000$, $T_{1e} = 1$ s, $T_{\text{ISS}} = 0.001$ s for different nuclear intrinsic relaxation times: $T_{1n}\rightarrow\infty$ (blue line), $T_{1n} = 10000$ s (red line), $T_{1n} = 1000$ s (gray line), $T_{1n} = 100$ s (orange line).}
\label{figure3}
\end{figure}

In \figurename~\ref{figure3} the effect of nuclear leakage is investigated. The thermal contact between the nuclear spin system and the lattice significantly affects the build-up curve even when
$T_{1n}$ is much slower than any other transition considered. For  $T_{1n} = 10000$ s, the final nuclear polarization goes down to  $P_n = 0.707$, which corresponds to a reduction of $15 \%$ with respect
to  $P_n$ in absence of leakage, while the corresponding polarization time becomes $T_{\text{pol}} = 910$ s.
\subsection{Competing $ISS$ and electron spin lattice regime}
The case of finite contact between electrons and nuclei, $T_{\text{ISS}} \approx T_{1e}$, is now considered.
\begin{figure}[t]
 \includegraphics[width=8.6cm]{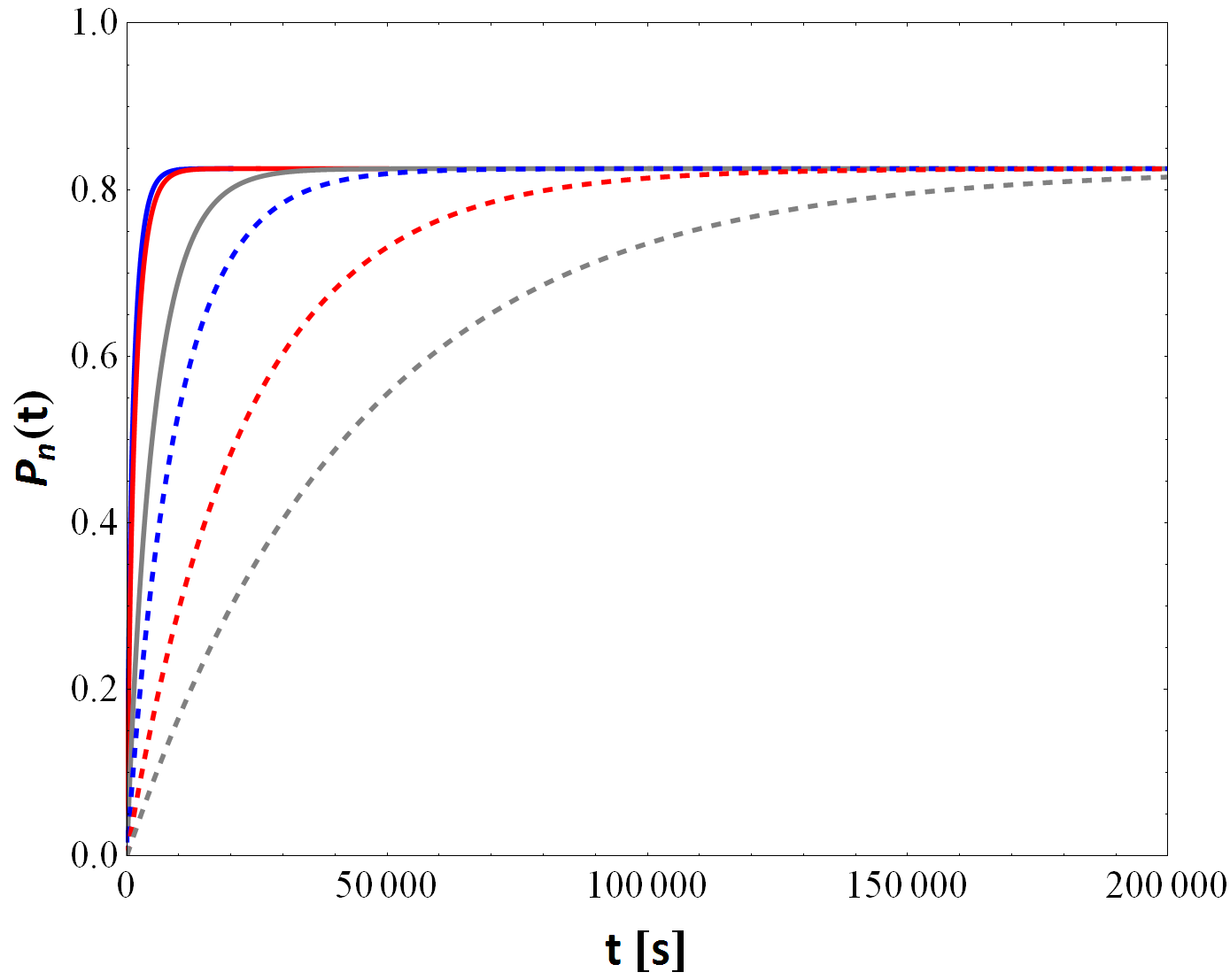}
\caption{Polarization build-up curves at 3.35 T and 1.2 K of a typical DNP system with $N_n/N_e = 1000$, $T_{1e} = 1$ s, $T_{1n} \rightarrow \infty $ for different values of contact time $T_{\text{ISS}}$ ($0.001$ s (solid blue line), $0.01$ s (solid red line), $0.1$ s (solid gray line), $0.2$ s (dashed blue line), $0.5$ s (dashed red line), $1$ s (dashed gray line)).}
\label{figure4a}
\end{figure} 

\begin{figure}[t]
 \includegraphics[width=8.6cm]{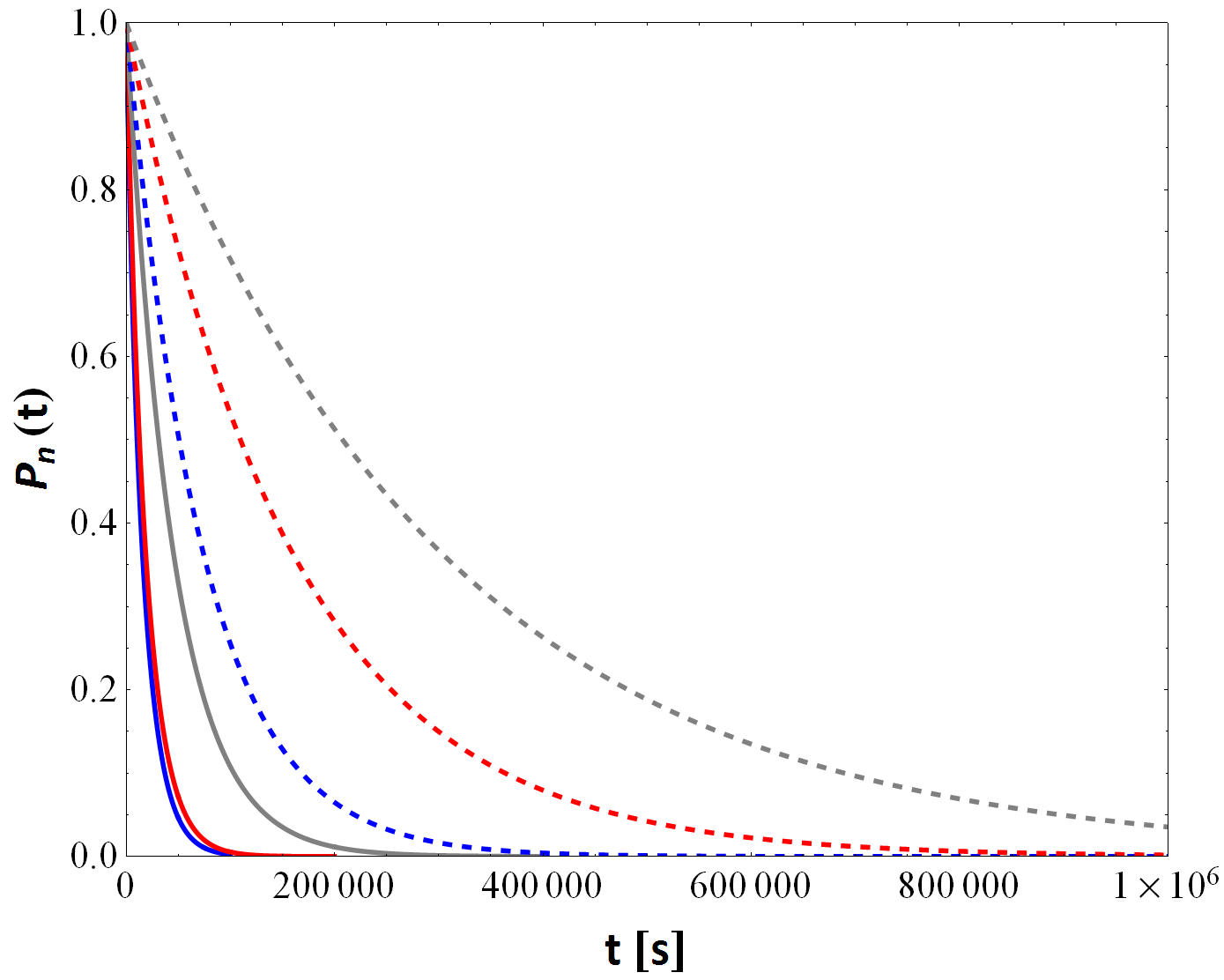}
\caption{Relaxation curves at 3.35 T and 1.2 K of a DNP system with $N_n/N_e = 1000$, $T_{1e} = 1$ s, $T_{1n} \rightarrow\infty$ for different values of contact time $T_{\text{ISS}}$ ($0.001$ s (solid blue line), $0.01$ s (solid red line), $0.1$ s (solid gray line), $0.2$ s (dashed blue line), $0.5$ s (dashed red line), $1$ s (dashed gray line)).}
\label{figure4b}
\end{figure}

In \figurename~\ref{figure4a} and \figurename~\ref{figure4b} build-up and relaxation curves, in absence of leakage ($T_{1n} \rightarrow \infty $) for $T_{1e} = 1$ s and $T_{\text{ISS}}$ within the range $0.001$ - $1$ s, are analyzed.
Two regimes can be identified. For fast $ISS$ process both $T_{\text{pol}}$ and $T_{\text{relax}}$ are not affected by the particular value of $T_{\text{ISS}}$, since the bottleneck process is represented by the electron spin-lattice relaxation. Conversely, when $T_{\text{ISS}}$ grows the $ISS$ process becomes the rate determining step and both polarization and relaxation times are enhanced. With $T_{\text{ISS}} = 0.1$ s for instance, one obtains $T_{\text{pol}} = 5300$ s 
and $T_{\text{relax}} = 44900$ s. The ratio between the two characteristic times remains constant and close to $10$. Finally no influence on the final polarization $P_n = 0.825$ is observed on varying  $T_{\text{ISS}}$.  

\begin{figure}[b]
 \includegraphics[width=8.6cm]{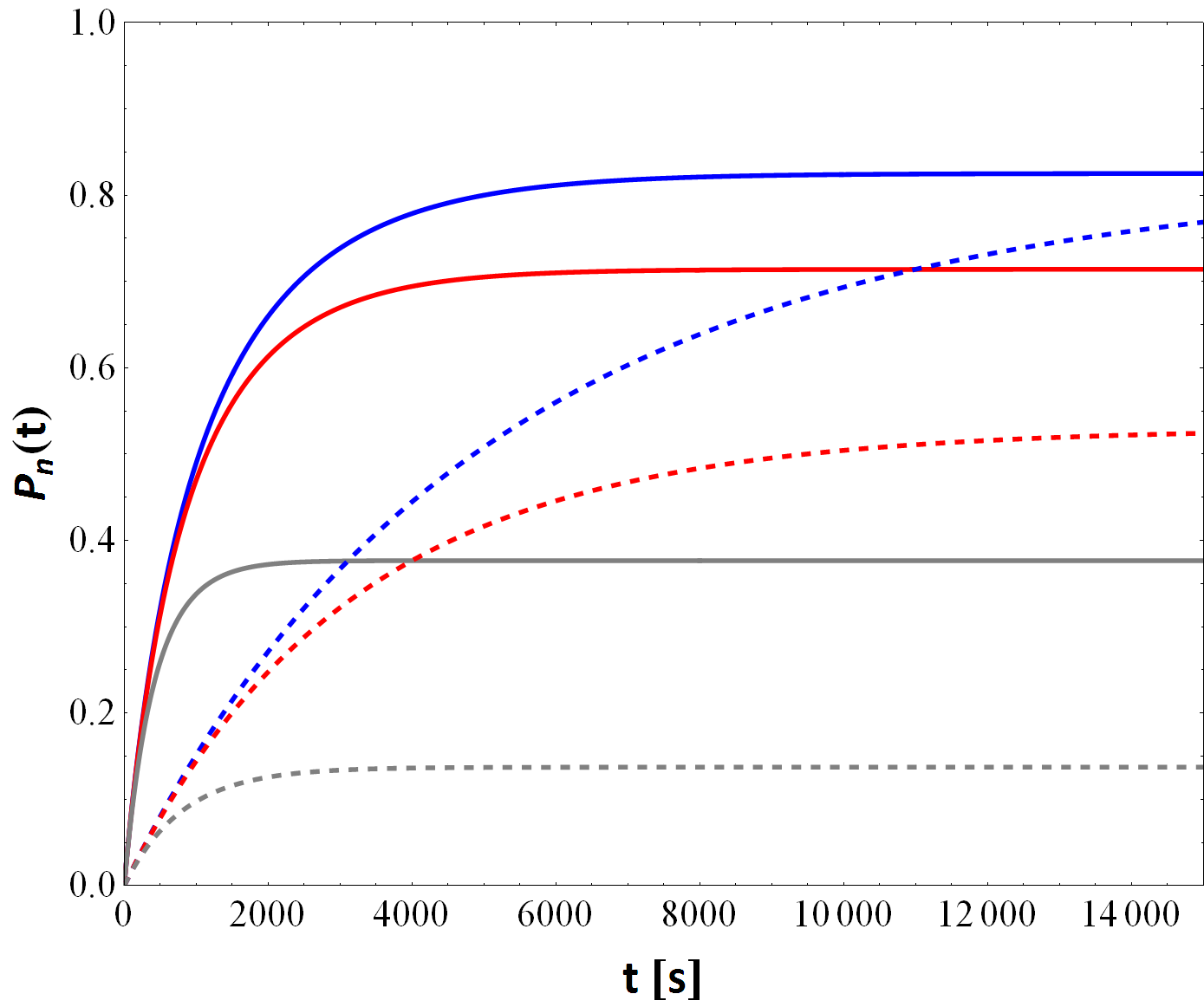}
\caption{Polarization build-up curves at 3.35 T and 1.2 K of a DNP system with $N_n/N_e = 1000$ and $T_{1e} = 1$ s for different values of contact time $T_{\text{ISS}}$ ($0.001$ s (solid lines) and $0.1$ s (dashed lines)) and nuclear intrinsic relaxation times $T_{1n}$ ($\rightarrow \infty$ (blue lines), $10000$ s (red lines) and $1000$ s (gray lines)).}
\label{figure5}
\end{figure}

Surprisingly, when an almost negligible leakage term is introduced  ($T_{1n} = 10000$ s), the final nuclear polarization becomes strongly dependent on the effectiveness  of the contact between nuclear and electron systems  (\figurename~\ref{figure5}). For  $T_{\text{ISS}} = 0.1$ s, the final nuclear polarization is $P_{n} = 0.529$, which corresponds to an important reduction of $36\%$ that could be relevant for explaining some experimental observations. The polarization time is measured equal to $3200$ s.
\begin{figure}[t]
 \includegraphics[width=8.2cm]{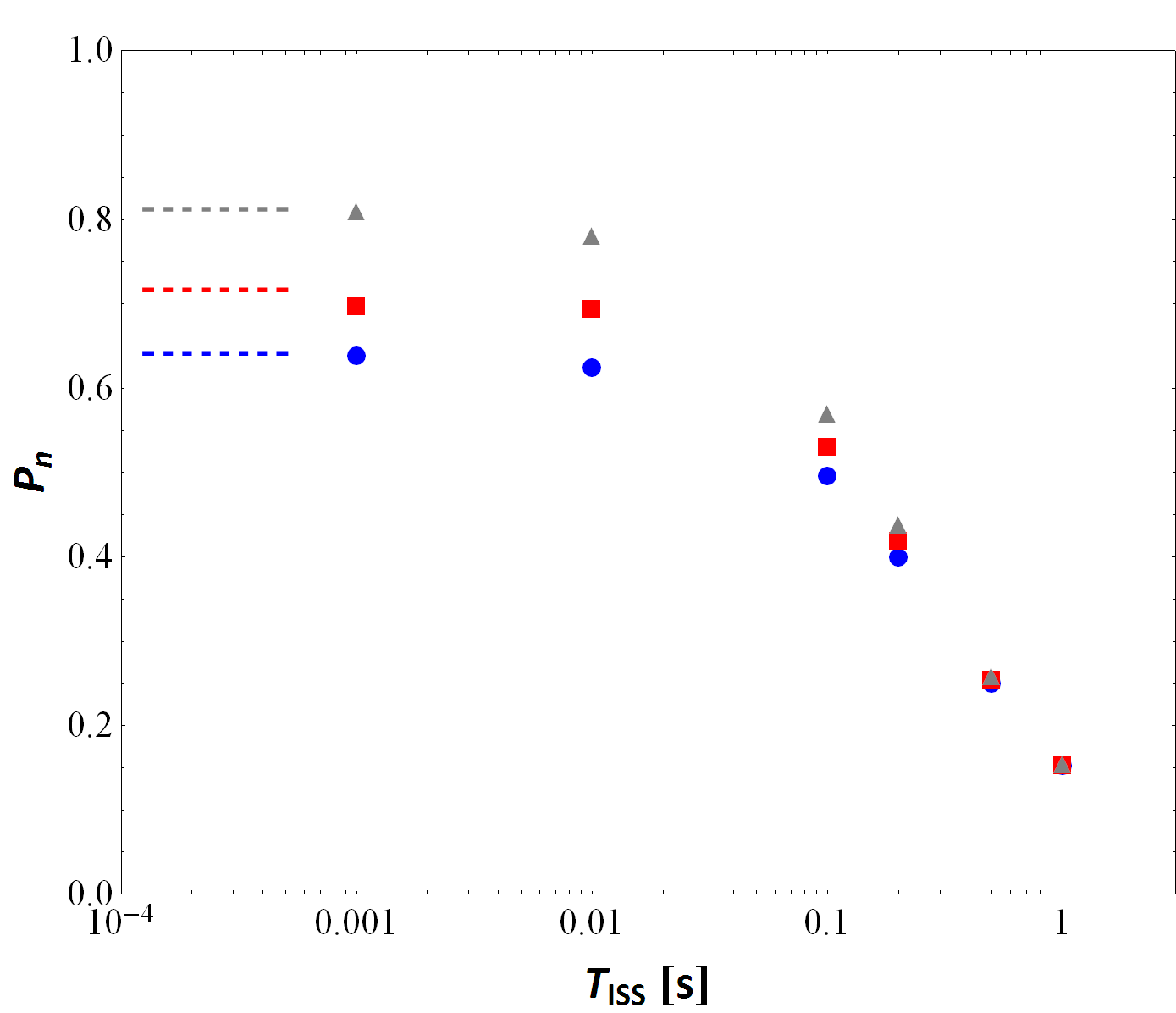}
\caption{Final polarization $versus$ contact time $T_{\text{ISS}}$ for different values of the electron relaxation time $T_{1e}$ ($2$ s (circles), $1$ s (squares) and $100$ ms (triangles)). Remaining parameters are set as follows: $N_n/N_e = 1000$, $T_{1n} = 10000$ s. Dashed lines represent polarization final values as estimated by the Borghini prediction in Eq.(\ref{BorghiniRelation}).}
\label{figure6a}
\end{figure} 

\begin{figure}[t]
 \includegraphics[width=8.6cm]{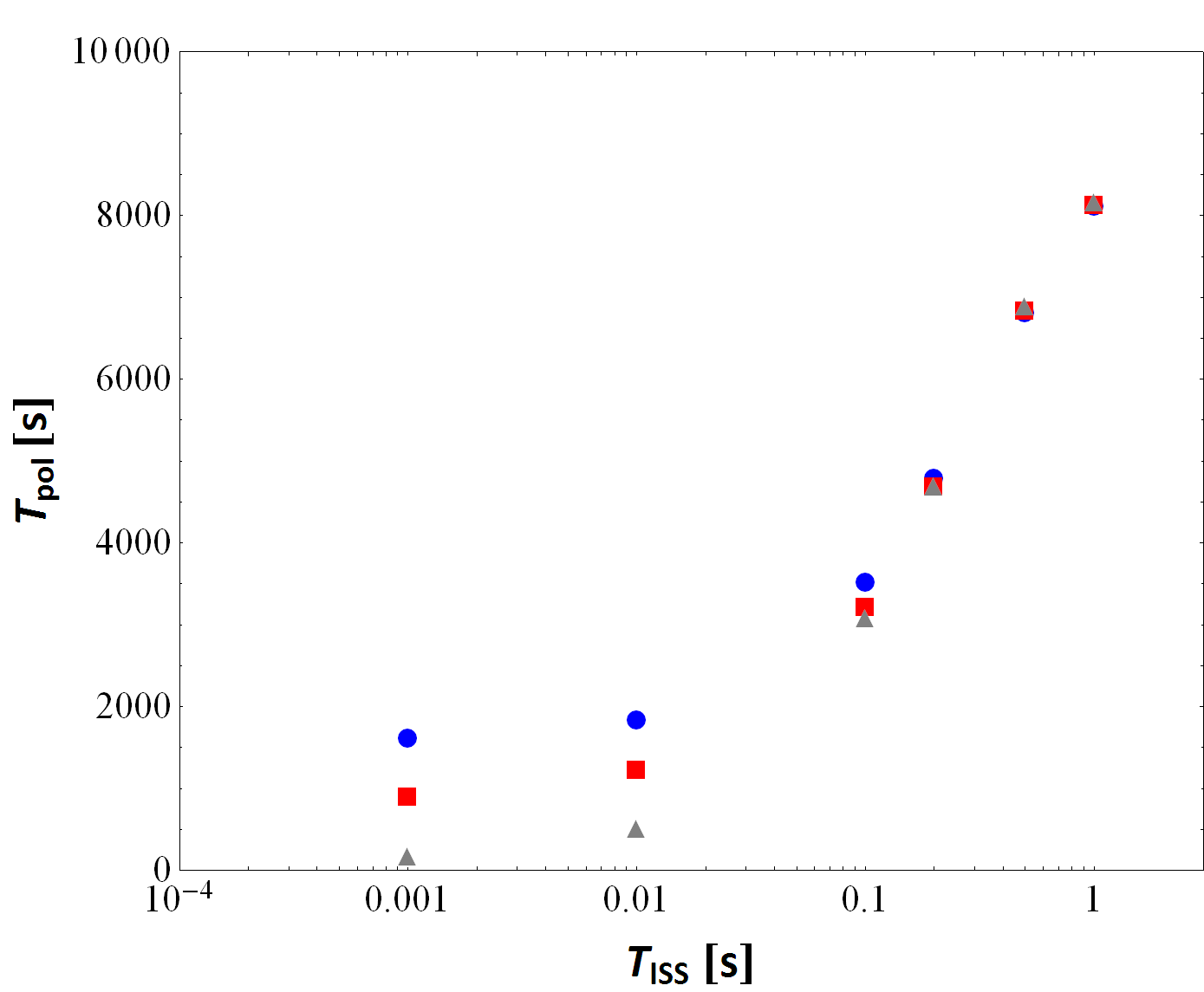}
\caption{Polarization build-up times as function of the contact time $T_{\text{ISS}}$ for different value of the electron relaxation time $T_{1e}$ ($2$ s (circles), $1$ s (squares) and $100$ ms (triangles)). Remaining parameters are set as follows: $N_n/N_e = 1000$, $T_{1n} = 10000$ s.}
\label{figure6b}
\end{figure}

A comprehensive summary of the role played by the different parameters is represented in \figurename~\ref{figure6a} and  \figurename~\ref{figure6b}.
Two distinct regimes are clearly observable: highly and poorly effective contact between nuclear and electron systems. The former ($T_{\text{ISS}} \rightarrow 0$)  is characterized by high levels of polarization and short build-up times, both dependent on $T_{1e}$, whilst the latter ($T_{\text{ISS}} \rightarrow \infty$) has low levels of polarization and long build-up times, both independent on $T_{1e}$. Clearly, in the first regime, the bottleneck role is played by $T_{1e}$, while $T_{\text{ISS}}$ is sufficiently high not to affect the system dynamics. In the second regime their role is reversed.

\section{Discussion and conclusion}
\label{Discussionandconclusion}
To date, the main attempt to give a theoretical description of the DNP phenomenon at low temperature in the TM regime is the prediction proposed by Borghini \cite{Borghini PRL, Abragam e Goldman} for the steady state nuclear polarization, 
that makes use of the standard parameterization:
\begin{eqnarray}
\label{congetturaPn}   
P_n &=& \tanh \left(\beta_n \omega_n \right) \quad \text{with} \, \beta_n = \hbar / (2 k_B T_n),
\end{eqnarray}
where $T_n$ represents the temperature reached by the nuclear reservoir under microwave irradiation. To compute this temperature, the original derivation conjectures that the final electron polarization profile ($P_{e, i} = P_{e, i} (t\rightarrow\infty)$) takes the form:
\begin{eqnarray}
\label{congetturaPe}   
P_{e, i} &=& - \tanh \left[ \beta (\Delta_i -c) \right].
\end{eqnarray}
Actually, when the spectral diffusion is the fastest process, this form can be rigorously derived - without any $ad$ $hoc$ conjecture - by imposing the detailed balance condition as shown in Appendix \ref{DetailedBalance}.
In addition to Eq.(\ref{congetturaPe}), a perfect thermal contact between the nuclear and the electron system and an infinite microwave power were assumed in the approach proposed by Borghini,  so that $\beta = \beta_n$  and $c = \Delta_0$.

Under these assumptions, the celebrated relation (re-derived in Appendix \ref{Borghirel} for convenience of the reader):
\begin{equation}
\label{BorghiniRelation}
\sum f_i (\Delta_i - \Delta_0) P_{e, i}+ \Delta_0 P_0 - \omega_n \frac{N_n T_{1e}}{N_{e}T_{1n}}P_n = 0
\end{equation}
gives a unique solution for $T_n$ and thus for $P_n$.
In the absence of leakage, the value of $P_n$ predicted by Eq.(\ref{BorghiniRelation}) at given temperature and field depends only on the ESR line shape and the irradiation frequency. For a Gaussian shape the maximal enhancement corresponds to $\omega_0 = \omega_e - \sigma$, where $\sigma$ is the standard deviation of the electron frequency distribution.
The Borghini prediction is in good qualitative agreement with many experimental observations obtained in low temperature DNP experiments. However it does not provide an accurate quantitative description of experimental data and in particular of
the maximum final nuclear polarization that turns always out to be overestimated \cite{Comment1, Comment2, Lumata2012, JHAL2008}.
When a typical  trityl doped sample, with an EPR line width of $63$ MHz, $T_{1e} = 1$ s, and $N_n/N_e =1000$ is considered, Eq.(\ref{BorghiniRelation}) predicts a maximum steady state polarization $P_n = 0.825$ in absence of leakage. The same
equilibrium value is obtained by our mathematical approach (\figurename~\ref{figure1} and  \figurename~\ref{figure4a}) under the same physical constrains.  

On the other hand the polarization levels achieved experimentally are in general much lower (see for instance Ref. \cite{JHAL2008, JHAL2010} and \cite{CMMI}).
To justify this discrepancy, one can call on dissipation terms. In fact, if a finite $T_{1n}$  is considered in Eq.(\ref{BorghiniRelation}), a loss of $P_n$ is obtained, that however, with $T_{1n}$ generally being longer
than  $10^4$ s \cite{JHAL2008}, can hardly exceed  $10$ - $15\%$ (cfr. \figurename~\ref{figure3}). A severe reduction of $P_n$ is obtained instead by means of the equation set (\ref{rateeq}), 
introducing a finite contact between nuclei and electrons in presence of a small leakage (see \figurename~\ref{figure5}). It is worth to notice that in absence of leakage,
{\em i.e.} when nuclei are isolated by the lattice, the final polarization is not affected by the efficiency of the contact between electron and nuclear systems.

Besides providing a more flexible scenario for $P_n$, the proposed mathematical framework allows the computation of polarization and relaxation times. Two regimes have been identified. 

When the contact between nuclei and electrons is highly efficient, the bottleneck of the spin dynamics is $T_{1e}$ and, by setting the parameters of the simulation according 
the experimental conditions used in \cite{JHAL2008} one obtains $T_{\text{pol}} \approx 10^3$ s and $T_{\text{relax}} \approx 10^4$ s, as effectively measured
in \cite{JHAL2008}.
It is interesting to observe that, in this fast exchange limit, an estimation of the order of magnitude of  $T_{\text{pol}}$ and $T_{\text{relax}}$  can be heuristically
derived as follows. During polarization, the nuclear system transfers energy to electrons, which are cooled by the lattice. 
In a time $T_{1e}$, the lattice can absorb an energy proportional at most to $N_e$, so that $T_{\text{pol}}\approx N_n T_{1e}/N_e \approx 10^3$ s, with $N_n/N_e = 1000$.
During relaxation, the effective number of electrons that can absorb energy from the nuclear system, proportional to  $(1-P_{e, i})(1+P_{e, i}) \approx (1-P_0^2)$,
collapses to $\approx 0.1$ due to the high electron polarization   $P_{e, i} \approx P_0 = 0.95$ at 3.35 T and 1.2 K, thus explaining the factor 
$T_{\text{relax}}/T_{\text{pol}}\approx 10$ observed both in simulations and experiments.

On decreasing the effectiveness of the contact between nuclear and electron spins, a different regime is established, where the $ISS$ process becomes the rate determining step  
and both $T_{\text{pol}}$ and $T_{\text{relax}}$ become longer. This provides a possible explanation of why  $T_{\text{pol}}$ can range between very different values in samples with
the same ratio  $N_e/N_n$ polarized in analogous conditions. By way of example, one can compare  the $[1$-$^{13}$C$]$-pyruvic acid samples doped with 15 mM 
of trityl radical studied in \cite{JHAL2008, JHAL2010}, where  $N_e/N_n = 1000$ and  $T_{\text{pol}} = 1200$ s with the  $[1$-$^{13}$C$]$-butyric acid sample mixed with $20\%$ in volume of DMSO and doped with 10 mM
trityl radical analyzed in \cite{CMMI}, having a much longer $T_{\text{pol}} = 3400$ s while the ratio $N_e/N_n $ remains almost unchanged.

In conclusion, we propose a novel approach based on rate equations for studying the dependency of dynamic nuclear polarization in the low temperature thermal mixing regime 
($T_{2e} \ll T_{1e}$) from the microscopic transitions involving electron and nuclear spins. This approach allows the recovery of the whole build-up curve and, in the limit of perfect 
contact between nuclei and electrons and infinite microwave power, leads to the same final nuclear polarization predicted by Borghini. In addition, by tuning the efficiency of the exchange 
interaction between nuclei and electrons, different values of $P_n$ are reached, providing an interpretation key for those experimental observations of $P_n$
which are not simply accounted for by leakage terms depending only by spin concentration and spin-lattice relaxation times.

The rate equation approach can be easily extended to more complex experimental systems. A second nuclear reservoir which also participates in TM could {\em e.g.} be included to interpret the dynamic experimental data measured in nitroxyl doped samples (where  both $^{13}$C and $^{1}$H Larmor 
frequencies do not exceed the ESR line width \cite{Comment1, Comment2, Sami}) or, similarly, in trityl doped samples containing $^{13}$C and $^{89}$Y nuclei,
both in contact with the electron reservoir \cite{Texas, Lumata2012}. Finally the versatility of the approach proposed here would easily allow 
to introduce new dissipative processes violating the precise assumptions and conservation principles the Borghini prediction is based on and that could possibly be useful to justify the many unexplained observations of low temperature DNP, \emph{e.g.} the reduction of 
$P_n$ on increasing $N_e$ \cite{JHAL2010, JHALhighfield}.

\section{Acknowledgement}
\label{Acknowledgement}
Christophe Texier and Thomas Guedr\'e are gratefully acknowledged for the fruitful discussions. This study has been supported in part by Regione Piemonte (POR FESR 2007/2013, line I.1.1), by the COST Action TD1103 (European Network for Hyperpolarization Physics and Methodology in NMR and MRI) and by ANR grant 09-BLAN-0097-02.
 
\appendix

\section{Detailed balance} \label{DetailedBalance}
After defining  $P_{e, i}^+$ as the fraction of electrons $up$ and $P_{e,i}^-$ as the fraction of electrons $down$ belonging to the packet $i$,
the detailed balance condition under the process depicted in \figurename~\ref{figureTransitions}, panel B, writes:
\begin{equation}\label{DB1}
P_{e, i-\delta}^+ \, \left(P_{e, i}^-\right)^2 \, P_{e, i+\delta}^+ = P_{e, i-\delta}^- \, \left(P_{e, i}^+\right)^2 \, P_{e, i+\delta}^-. 
\end{equation}
Then, by using the relation 
\begin{eqnarray}
P_{e, i}^+=\frac{(1+P_{e, i})}{2}  \nonumber \\
P_{e, i}^-=\frac{(1-P_{e, i})}{2}  \nonumber 
\end{eqnarray}
one comes to an equation for the electron polarization
\begin{eqnarray}\label{DB2}
\left(1+P_{e, i-\delta}\right) \,\left(1-P_{e, i}\right)^2 \, \left(1+P_{e, i+\delta}\right) =  \nonumber \\
=\left(1-P_{e, i-\delta}\right) \,\left(1+P_{e, i}\right)^2 \, \left(1-P_{e, i+\delta}\right), 
\end{eqnarray}
that can be solved in the continuum limit where $\delta \omega \rightarrow 0$, $N_p \rightarrow \infty$ and $P_{e, i} \rightarrow P_e(\Delta_i)$. 
It is sufficient to consider the transitions between consecutive packets, so that $P_{e, i + \delta} = P_{e, i+1}\rightarrow P_e(\Delta_i+\delta \omega)$, and write a second order expansion
\begin{equation}\label{DBx}
P_e(\Delta_i+ \delta \omega) \approx P_e(\Delta_i)+ \delta \omega P'_e (\Delta_i) +\frac{\delta \omega^2}{2} P''_e (\Delta_i).
\end{equation}
By combining relations  (\ref{DB2}) and (\ref{DBx}), one gets the second order differential equation 
\begin{equation}\label{DB3}
2 P_e(\Delta_i) P'_e(\Delta_i)^2+  P''_e(\Delta_i) \left(1 - P_e(\Delta_i)^2\right)=0,
\end{equation}
whose general solution
\begin{equation}\label{DB4}
P_e(\Delta_i) =-\tanh\left( \beta(\Delta_i-c)\right)
\end{equation}
was used in our treatment as starting point for what we called ``STT'' process. The same parametric function has been postulated in the derivation of the Borghini prediction for the steady state.

\section{Microscopic derivation of rate equations} \label{RateEqDerivation}
The term proportional to $1/T_{\text{ISS}}$ and describing the $ISS$ process, used in the rate equations (\ref{rateeq}), is here derived for the electron polarization first and then for  nuclear polarization.

\subsection{Electron polarization}
\begin{figure}[t]
 \includegraphics[width=8cm]{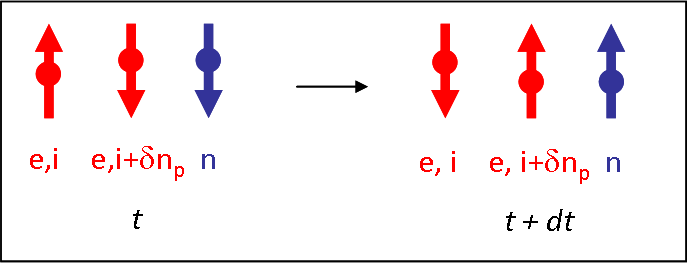}
\caption{Schematic representation of one possible $ISS$ event.}
\label{figureISS}
\end{figure}

Be $P_{e, i}^+$ the fraction of electrons $up$ belonging to the packet $i$, $P_{n}^+$ the fraction of nuclei $up$, $P_{e,i}^-$ the fraction of electrons $down$ belonging to the packet $i$ and $P_{n}^-$ the fraction of nuclei $down$.
When the $ISS$ event depicted in \figurename~\ref{figureISS} occurs, the fraction of electrons $up$ in the $i$-th packet is reduced by $1/(N_e f_i)$.
The number of possible transitions is the product of:
\begin{itemize}
	\item the number of the electrons $up$ in the $i$-th packet: $N_e f_i P_{e,i}^+$,
	\item the number of the electrons $down$ in the $(i + \delta n_p)$-th packet: $N_e f_{i+\delta n_p} P_{e, i + \delta n_p}^-$,
	\item the number of nuclei $dow$n: $N_n P_n^-$. 
\end{itemize}
The rate of such process is $1/ (T_{\text{ISS}} N_e N_n)$, and the total reduction of $P_{e, i}^+$ in the time interval $dt$ is:
\begin{equation} 
-\frac{dt}{T_{\text{ISS}}} f_{i+\delta n_p} P_{e,i}^+ P_{e, i + \delta n_p}^- P_n^-. 
\end{equation}
The total variation of $P_{e, i}^+$ induced by all possible $ISS$ transitions, $\delta P_{e, i}^+$, is given by:
\begin{eqnarray} 
\delta P_{e, i}^+ &=& \frac{dt}{T_{\text{ISS}}} \left[ f_{i+\delta n_p} \left( P_{e,i}^- P_{e, i + \delta n_p}^+ P_n^+ - P_{e,i}^+ P_{e, i + \delta n_p}^- P_n^-\right) \right. \nn \nonumber \\
&& \left. \small{+} f_{i-\delta n_p} \left( P_{e,i}^- P_{e, i - \delta n_p}^+ P_n^- - P_{e,i}^+ P_{e, i - \delta n_p}^- P_n^+\right)\right]\nonumber. 
\end{eqnarray}
Using the relations: 
\begin{eqnarray}
P_{e, i}^+=\frac{(1+P_{e, i})}{2}, \quad
P_{e, i}^-=\frac{(1-P_{e, i})}{2}  \nonumber \\
P_{n}^+=\frac{(1+P_{n})}{2}, \quad
P_{n}^-=\frac{(1-P_{n})}{2},  \nonumber 
\end{eqnarray}
the total variation of $P_{e, i}$ induced by all possible $ISS$ processes, $\delta P_{e, i} =2 \delta P_{e, i}^+$, can be written as follows:
\begin{eqnarray} 
\label{RE1}
\delta P_{e, i} &=& \frac{dt}{4 T_{\text{ISS}}} \left\{ f_{i+\delta n_p} \left[ (1\small{-}P_{e,i}) (1\small{+}P_{e, i + \delta n_p}) (1\small{+}P_n) \right. \right. \nn \nonumber \\
&& \left. - (1\small{+}P_{e,i})(1\small{-}P_{e, i + \delta n_p})(1\small{-}P_n)\right] \nonumber \\
&& \small{+} f_{i-\delta n_p} \left[ (1\small{-}P_{e,i})(1\small{+}P_{e, i - \delta n_p})(1\small{-}P_n) \right. \nonumber \\
&& \left. \left. -(1\small{+}P_{e,i})(1\small{-}P_{e, i - \delta n_p})(1\small{+}P_n)\right] \right\}. 
\end{eqnarray}
The term proportional to $1/T_{\text{ISS}}$ in the first equation of set (\ref{rateeq}) can be now easily derived from Eq.(\ref{RE1}) by means of simple algebraic calculations.

\subsection{Nuclear polarization}
When the event depicted in \figurename~\ref{figureISS} occurs, the  fraction of nuclei $up$ is increased of a factor $1/N_n$.
The number of possible transitions is the product of:
\begin{itemize}
	\item the number of the electrons $up$ in the $i$-th packet: $N_e f_i P_{e,i}^+$,
	\item the number of the electrons $down$ in the $(i + \delta n_p)$-th packet: $N_e f_{i+\delta n_p} P_{e, i + \delta n_p}^-$,
	\item the number of nuclei $down$: $N_n P_n^-$. 
\end{itemize}
The rate of such process is $1/ (T_{\text{ISS}} N_e N_n)$, and the relevant increment of $P_{n}^+$ in the time interval $dt$:
\begin{equation} 
\frac{N_e dt}{N_n T_{\text{ISS}}} f_i f_{i+\delta n_p} P_{e,i}^+ P_{e, i + \delta n_p}^- P_n^-. 
\end{equation}
Considering now all the possible processes, the total variation of $P_{n}^+$ induced by the $ISS$ process, $\delta P_{n}^+$, is given by:
\begin{eqnarray} 
\delta P_{n}^+ &\small{=}& \frac{N_e dt}{N_n T_{\text{ISS}}} \sum_i f_i f_{i+\delta n_p} \left[ P_{e,i}^+ P_{e, i + \delta n_p}^- P_n^- \right. \nonumber \\
&& \left. \small{-} P_{e,i}^- P_{e, i + \delta n_p}^+ P_n^+ \right]. 
\end{eqnarray}
Following the line previously described for $P_{e, i}$, one immediately arrives to the equations for $P_n$ reported in the main text (\ref{rateeq}).

\section{Borghini relation} \label{Borghirel}
To facilitate the reading of the manuscript, we report the derivation of the Borghini relation in Eq.(\ref{BorghiniRelation}) according to the line proposed in \cite{Abragam e Goldman}.
The energy of the whole electron and nuclear system is conveniently split into two reservoirs: the Zeeman electron contribution 
\begin{equation}
E_{Z_e}(t) = \frac{1}{2} N_e \hbar \omega_e \sum f_i P_{e, i}(t)
\end{equation}
and the non-Zeeman electron plus Zeeman nuclear term
\begin{equation}
E_{\text{NZ}-Z_n}(t) = \frac{1}{2} N_n \hbar \omega_n P_n(t) - \frac{1}{2} N_e \hbar \sum \Delta_i f_i P_{e, i}(t).
\end{equation}

The time evolution of the two energy reservoirs is described by the following equations:
\begin{eqnarray}
\label{dEdt}
\frac{d E_{Z_e}(t)}{dt} &=& \frac{1}{2} N_e \hbar \omega_e \sum f_i \frac{dP_{e, i}(t)}{dt} \\
\frac{dE_{\text{NZ}-Z_n}(t)}{dt} &=& \frac{1}{2} N_n \hbar \omega_n \frac{dP_n(t)}{dt} - \frac{1}{2} N_e \hbar \sum \Delta_i f_i \frac{dP_{e, i}(t)}{dt},\nonumber
\end{eqnarray}
and since either the spectral diffusion and the $ISS$ process conserve both $E_{Z_e}(t)$ and $E_{\text{NZ}-Z_n}(t)$, one obtains:
\begin{eqnarray}
\label{dEdt2}
\frac{d E_{Z_e}(t)}{dt} &=& \frac{1}{2} N_e \hbar \omega_e \left[ \sum f_i \frac{P_{0}-P_{e, i}(t)}{T_{1e}} - f_0 \frac{P_{e, 0}}{T_{1 \text{MW}}}\right ]\nonumber\\
\frac{dE_{\text{NZ}-Z_n}(t)}{dt} &=& \frac{1}{2} N_e \hbar \left[ \sum \Delta_i f_i \frac{P_{e, i}(t)}{T_{1e}}  + f_0 \Delta_0 \frac{P_{e, 0}}{T_{1 \text{MW}}} \right ]\nonumber \\
&&- \frac{1}{2} N_n \hbar \omega_n \frac{P_n(t)}{T_{1n}}.
\end{eqnarray}
It is important to observe that the evolution of the two energy reservoirs depends on the time progression of all $P_{e, i}(t)$ and $P_n(t)$, that is the full solution of the system of rate equations reported in Eq.(\ref{rateeq}). 
As far as only the steady state solution is required, however, it is sufficient to impose the simultaneous vanishing of both right-hand sides of Eq.(\ref{dEdt2}). Thus, by multiplying the first Eq.(\ref{dEdt2}) by $\Delta_0 / \omega_e$ and adding it to the right-hand side of the second Eq.(\ref{dEdt2}), one gets rid of the microwave transition probability $1/T_{1\text{MW}}$ and obtains the celebrated Borghini relation given in Eq.(\ref{BorghiniRelation}).

\end{document}